# *101 Formulaic Alphas*

Zura Kakushadze[§†1]

[§] *Quantigic® Solutions LLC,[2] 1127 High Ridge Road, #135, Stamford, CT 06905*

[†] *Free University of Tbilisi, Business School & School of Physics
240, David Agmashenebeli Alley, Tbilisi, 0159, Georgia*

December 9, 2015

*"There are two kinds of people in this world:
Those seeking happiness, and bullfighters."*
(Zura Kakushadze, ca. early '90s)[3]

Abstract

We present explicit formulas – that are also computer code – for 101 real-life quantitative trading alphas. Their average holding period approximately ranges 0.6-6.4 days. The average pair-wise correlation of these alphas is low, 15.9%. The returns are strongly correlated with volatility, but have no significant dependence on turnover, directly confirming an earlier result based on a more indirect empirical analysis. We further find empirically that turnover has poor explanatory power for alpha correlations.

---

[1] Zura Kakushadze, Ph.D., is the President and a Co-Founder of Quantigic® Solutions LLC and a Full Professor in the Business School and the School of Physics at Free University of Tbilisi. Email: zura@quantigic.com

[2] DISCLAIMER: This address is used by the corresponding author for no purpose other than to indicate his professional affiliation as is customary in publications. In particular, the contents of this paper are not intended as an investment, legal, tax or any other such advice, and in no way represent views of Quantigic® Solutions LLC, the website www.quantigic.com or any of their other affiliates.

[3] Paraphrasing *Blondie's* (Clint Eastwood) one-liners from a great 1966 motion picture *The Good, the Bad and the Ugly* (directed by Sergio Leone).

## 1. Introduction

There are two complementary – and in some sense even competing – trends in modern quantitative trading. On the one hand, more and more market participants (e.g., quantitative traders, inter alia) employ sophisticated quantitative techniques to mine alphas.[4] This results in ever fainter and more ephemeral alphas. On the other hand, technological advances allow to essentially automate (much of) the alpha harvesting process. This yields an ever increasing number of alphas, whose count can be in hundreds of thousands and even millions, and with the exponentially increasing progress in this field will likely be in billions before we know it…

This proliferation of alphas – albeit mostly faint and ephemeral – allows combining them in a sophisticated fashion to arrive at a unified "mega-alpha". It is then this "mega-alpha" that is actually traded – as opposed to trading individual alphas – with a bonus of automatic internal crossing of trades (and thereby crucial-for-profitability savings on trading costs, etc.), alpha portfolio diversification (which hedges against any subset of alphas going bust in any given time period), and so on. One of the challenges in combining alphas is the usual "too many variables, too few observations" dilemma. Thus, the alpha sample covariance matrix is badly singular.

Also, naturally, quantitative trading is a secretive field and data and other information from practitioners is not readily available. This inadvertently creates an enigma around modern quant trading. E.g., with such a large number of alphas, are they not highly correlated with each other? What do these alphas look like? Are they mostly based on price and volume data, mean-reversion, momentum, etc.? How do alpha returns depend on volatility, turnover, etc.?

In a previous paper [Kakushadze and Tulchinsky, 2015] took a step in demystifying the realm of modern quantitative trading by studying some empirical properties of 4,000 real-life alphas. In this paper we take another step and present explicit formulas – that are also computer code – for 101 real-life quant trading alphas. Our formulaic alphas – albeit most are not necessarily all that "simple" – serve a purpose of giving the reader a glimpse into what some of the simpler real-life alphas look like.[5] It also enables the reader to replicate and test these alphas on historical data and do new research and other empirical analyses. Hopefully, it further inspires (young) researchers to come up with new ideas and create their own alphas.

---

[4] "An alpha is a combination of mathematical expressions, computer source code, and configuration parameters that can be used, in combination with historical data, to make predictions about future movements of various financial instruments." [Tulchinsky *et al*, 2015] Here "alpha" – following the common trader lingo – generally means any reasonable "expected return" that one may wish to trade on and is not necessarily the same as the "academic" alpha. In practice, often the detailed information about how alphas are constructed may even not be available, e.g., the only data available could be the position data, so "alpha" then is a set of instructions to achieve certain stock (or other instrument) holdings by some times $t_1$, $t_2$, … (e.g., a tickers by holdings matrix for each $t_s$).

[5] We picked these alphas largely based on simplicity considerations, so they can be presented within the inherent limitations of a paper. There also exist myriad other, "non-formulaic" (coded and too-complex-to-present) alphas.

We discuss some general features of our formulaic alphas in Section 2. These alphas are mostly "price-volume" (daily close-to-close returns, open, close, high, low, volume and vwap) based, albeit "fundamental" input is used in some of the alphas, including one alpha utilizing market cap, and a number of alphas employing some kind of a binary industry classification such as GICS, BICS, NAICS, SIC, etc., which are used to industry-neutralize various quantities.[6]

We discuss empirical properties of our alphas in Section 3 based on data for individual alpha Sharpe ratio, turnover and cents-per-share, and also on a sample covariance matrix. The average holding period approximately ranges from 0.6 to 6.4 days. The average (median) pair-wise correlation of these alphas is low, 15.9% (14.3%). The returns $R$ are strongly correlated with the volatility $V$, and as in [Kakushadze and Tulchinsky, 2015] we find an empirical scaling

$$R \sim V^X \qquad (1)$$

with $X \approx 0.76$ for our 101 alphas. Furthermore, we find that the returns have no significant dependence on the turnover $T$. This is a direct confirmation of an earlier result by [Kakushadze and Tulchinsky, 2015], which is based on a more indirect empirical analysis.[7]

We further find empirically that the turnover per se has poor explanatory power for alpha correlations. This is not to say that the turnover does not add value in, e.g., modeling the covariance matrix via a factor model.[8] A more precise statement is that pair-wise correlations $\Psi_{ij}$ of the alphas ($i, j = 1, \ldots, N$ label the $N$ alphas, $i \neq j$) are not highly correlated with the product $\ln(\tau_i)\ln(\tau_j)$, where $\tau_i = T_i / \mu$, and $\mu$ is an a priori arbitrary normalization constant.[9]

We briefly conclude in Section 4. Appendix A contains our formulaic alphas with definitions of the functions, operators and input data used therein. Appendix B contains some legalese.

## 2. Formulaic Alphas

In this section we describe some general features of our 101 formulaic alphas. The alphas are proprietary to WorldQuant LLC and are used here with its express permission. We provide as many details as we possibly can within the constraints imposed by the proprietary nature of the alphas. The formulaic expressions – that are also computer code – are given in Appendix A.

---

[6] More precisely, depending on the alpha and industry classification used, neutralization can be w.r.t. sectors, industries, subindustries, etc. – different classifications use different nomenclature for levels of similar granularity.

[7] In [Kakushadze and Tulchinsky, 2015] the alpha return volatility was not directly available and was estimated indirectly based on the Sharpe ratio, cents-per-share and turnover data. Here we use direct realized volatility data.

[8] Depending on a construction, a priori the turnover might add value via the specific (idiosyncratic) risk for alphas.

[9] Here we use log of the turnover as opposed to the turnover itself as the latter has a skewed, roughly log-normal distribution, while pair-wise correlations take values in $(-1, 1)$ (in fact, their distribution is tighter – see below).

Very coarsely, one can think of alpha signals as based on mean-reversion or momentum.[10] A mean-reversion alpha has a sign opposite to the return on which it is based. E.g., a simple mean-reversion alpha is given by

$$-\ln(\text{today's open} / \text{yesterday's close}) \qquad (2)$$

Here yesterday's close is adjusted for any splits and dividends if the ex-date is today. The idea (or hope) here is that the stock will mean-revert and give back part of the gains (if today's open is higher than yesterday's close) or recoup part of the losses (if today's open is lower than yesterday's close). This is a so-called "delay-0" alpha. Generally, "delay-0" means that the time of some data (e.g., a price) used in the alpha coincides with the time during which the alpha is intended to be traded. E.g., the alpha (2) would ideally be traded at or, more realistically, as close as possible to today's open. More broadly, this can be some other time, e.g., the close.[11]

A simple example of a momentum alpha is given by

$$\ln(\text{yesterday's close} / \text{yesterday's open}) \qquad (3)$$

Here it makes no difference if the prices are adjusted or not. The idea (or hope) here is that if the stock ran up (slid down) yesterday, the trend will continue today and the gains (losses) will be further increased. This is a so-called "delay-1" alpha if the intent is to trade it today (e.g., starting at the open).[12] Generally, "delay-1" means that the alpha is traded on the day subsequent to the date of the most recent data used in computing it. A "delay-$d$" alpha is defined similarly, with $d$ counting the number of days by which the data used is out-of-sample.

In complex alphas elements of mean-reversion and momentum can be mixed, making them less distinct in this regard. However, one can think of smaller building blocks of such alphas as being based on mean-reversion or momentum. For instance, Alpha#101 in Appendix A is a delay-1 momentum alpha: if the stock runs up intraday (i.e., close > open and high > low), the next day one takes a long position in the stock. On the other hand, Alpha#42 in Appendix A essentially is a delay-0 mean-reversion alpha: rank(vwap – close) is lower if a stock runs up in the second half of the day (close > vwap)[13] as opposed to sliding down (close < vwap). The denominator weights down richer stocks. The "contrarian" position is taken close to the close.

---

[10] On longer horizons, for a discussion of mean-reversion (contrarian) and momentum (trend following) strategies, see, e.g., [Avellanida and Lee, 2010] and [Jegadeesh and Titman, 1993], respectively, and references therein.

[11] Four of our 101 alphas in Appendix A, namely, the alphas numbered 42, 48, 53 and 54, are delay-0 alphas. They are assumed to be traded at or as close as possible to the close of the trading day for which they are computed.

[12] On the other hand, if the alpha (3) is executed as close as possible to yesterday's close, then it is delay-0.

[13] Here "vwap", as usual, stands for "volume-weighted average price".

## 3. Data and Empirical Properties of Alphas

In this section we describe empirical properties of our formulaic alphas based on data proprietary to WorldQuant LLC, which is used here with its express permission. We provide as many details as possible within the constraints of the proprietary nature of this dataset.

For our alphas we take the annualized daily Sharpe ratio $S$, daily turnover $T$, and cents-per-share $C$. Let us label our alphas by the index $i$ ($i = 1, ..., N$), where $N = 101$ is the number of alphas. For each alpha, $S_i$, $T_i$ and $C_i$ are defined via

$$S_i = \sqrt{252}\, \frac{P_i}{V_i} \tag{4}$$

$$T_i = \frac{D_i}{I_i} \tag{5}$$

$$C_i = 100\, \frac{P_i}{Q_i} \tag{6}$$

Here: $P_i$ is the average daily P&L (in dollars); $V_i$ is the daily portfolio volatility; $Q_i$ is the average daily shares traded (buys plus sells) by the $i$-th alpha; $D_i$ is the average daily dollar volume traded; and $I_i$ is the total dollar investment in said alpha (the actual long plus short positions, without leverage). More precisely, the principal of $I_i$ is constant; however, $I_i$ fluctuates due to the daily P&L. So, both $D_i$ and $I_i$ are adjusted accordingly (such that $I_i$ is constant) in Equation (4). The period of time over which this data is collected is Jan 4, 2010-Dec 31, 2013. For the same period we also take the sample covariance matrix $Y_{ij}$ of the realized daily returns for our alphas. The number of observations in the time series is 1,006, and $Y_{ij}$ is nonsingular. From $Y_{ij}$ we read off the daily return volatility $\sigma_i^2 = Y_{ii}$ and the correlation matrix $\Psi_{ij} = Y_{ij} / \sigma_i \sigma_j$ (where $\Psi_{ii} = 1$). Note that $V_i = \sigma_i I_i$, and the average[14] daily return is given by $R_i = P_i / I_i$.

Table 1 and Figure 1 summarize the data for the annualized Sharpe ratio $S_i$, daily turnover, $T_i$, average holding period $1 / T_i$, cents-per-share $C_i$, daily return volatility $\sigma_i$, annualized average daily return $\tilde{R}_i = 252\, R_i$, and $N(N-1)/2$ pair-wise correlations $\Psi_{ij}$ with $i > j$.

### 3.1. Return v. Volatility & Turnover

We run two cross-sectional regressions, both with the intercept, of $\ln(R_i)$ over i) $\ln(\sigma_i)$ as the sole explanatory variable, and ii) over $\ln(\sigma_i)$ and $\ln(T_i)$. The results are summarized in Tables 2 and 3. Consistently with [Kakushadze and Tulchinsky, 2015], we have no statistically significant dependence on the turnover $T_i$ here, while the average daily return $R_i$ is strongly correlated with the daily return volatility $\sigma_i$ and we have the scaling property (1) with $X \approx 0.76$.

---

[14] Here the average is over the time series of the realized daily returns.

### *3.2. Does Turnover Explain Correlations?*

If we draw a parallel between alphas and stocks, then alpha turnover is analogous to stock liquidity, which is typically measured via an average daily dollar volume (ADDV).[15] Log of ADDV is routinely used as a style risk factor[16] in multifactor risk models[17] for approximating stock portfolio covariance matrix structure, whose chief goal is to model the off-diagonal elements of the covariance matrix, that is, the pair-wise correlation structure.[18] Following this analogy, we can ask if the turnover – or more precisely its log – has explanatory power for modeling alpha correlations.[19] It is evident that using the turnover directly (as opposed to its log) would get us nowhere due to the highly skewed (roughly log-normal) turnover distribution (see Figure 1).

To answer this question, recall that in a factor model the covariance matrix is modeled via

$$\Gamma_{ij} = \xi_i^2 \, \delta_{ij} + \sum_{A,B=1}^{K} \Omega_{iA} \, \varphi_{AB} \, \Omega_{jB} \qquad (7)$$

Here: $\xi_i^2$ is the specific risk; $\Omega_{iA}$ is an $N \times K$ factor loadings matrix corresponding to $K \ll N$ risk factors; and $\varphi_{AB}$ is a factor covariance matrix. In our case, we are interested in modeling the correlation matrix $\Psi_{ij}$ and ascertaining whether the turnover has explanatory power for pair-wise correlations. Whether the volatility and turnover are correlated is a separate issue.

So, our approach is to take one of the columns of the factor loadings matrix as $\ln(T_i)$. More precisely, a priori there is no reason why we should pick $\ln(T_i)$ as opposed to $\ln(\tau_i)$, where $\tau_i = T_i / \mu$, and $\mu$ is some normalization factor. To deal with this, let us normalize $\tau_i$ such that $\ln(\tau_i)$ has zero cross-sectional mean, and let $\nu_i = 1$ be the unit $N$-vector (the intercept). Then we can construct three symmetric tensor combinations $x_{ij} = \nu_i \nu_j$, $y_{ij} = \nu_i \ln(\tau_j) + \nu_j \ln(\tau_i)$, and $z_{ij} = \ln(\tau_i) \ln(\tau_j)$. Let us now define a composite index $\{a\} = \{(i,j) | i > j\}$, which takes $M = N(N-1)/2$ values, i.e., we pull the off-diagonal lower-triangular elements of a general symmetric matrix $G_{ij}$ into a vector $G_a$. This way we can construct four $M$-vectors $\Psi_a$, $x_a$, $y_a$ and $z_a$. Now we can run a linear regression of $\Psi_a$ over $x_a$, $y_a$ and $z_a$. Note that $x_a = 1$ is simply the intercept (the unit $M$-vector), so this is a regression of $\Psi_a$ over $y_a$ and $z_a$ with the intercept. The results are summarized in Table 4. It is evident that the linear and bilinear (in $\ln(\tau_i)$) variables $y_a$ and $z_a$ have poor explanatory power for pair-wise correlations $\Psi_a$, while $x_a$ (the intercept) simply models the average correlation $\text{Mean}(\Psi_a)$. Recall that by construction $y_a$ and $z_a$ are orthogonal to $x_a$, and these three explanatory variables are independent of each other.

---

[15] Perhaps a more precise analogy would be between the turnover and the ratio of ADDV and market cap; however, this is not going to be critical for our purposes here.

[16] For liquidity as a style risk factor, see, e.g., [Pastor and Stambaugh, 2003] and references therein.

[17] See, e.g., [Grinold and Kahn, 2000] and references therein.

[18] Variances are relatively stable and can be computed based on historical data (sample variances). It is the off-diagonal elements of the sample covariance matrix – to wit, the correlations – that are out-of-sample unstable.

[19] Log of the turnover as a factor for risk models for alpha portfolios was suggested in [Kakushadze, 2014].

Let us emphasize that our conclusion does not necessarily mean the turnover adds no value in the factor model context, it only means that the turnover per se does not appear to help in modeling pair-wise alpha correlations. The above analysis does not address whether the turnover adds explanatory value to modeling variances, e.g., the specific risk.[20] Thus, a linear regression of $\ln(\sigma)$ over $\ln(T)$ (with the intercept) shows nonzero correlation between these variables (see Table 5), albeit not very strong. To see if the turnover adds value via, e.g., the specific risk requires using certain proprietary methods outside of the scope of this paper.[21]

## 4. Conclusions

We emphasize that the 101 alphas we present here are not "toy" alphas but real-life trading alphas used in production. In fact, 80 of these alphas are in production as of this writing.[22] To our knowledge, this is the first time such a large number of real-life explicit formulaic alphas appear in the literature. This should come as no surprise: naturally, quant trading is highly proprietary and secretive. Our goal here is to provide a glimpse into the complex world of modern and ever-evolving quantitative trading and help demystify it, to any degree possible.

Technological advances nowadays allow automation of alpha mining. Quantitative trading alphas are by far the most numerous of available trading signals that can be turned into trading strategies/portfolios. There are myriad permutations of individual stock holdings in a (dollar-neutral) portfolio of, e.g., 2,000 most liquid U.S. stocks that can result in a positive return on high- and mid-frequency time horizons. In addition, many of these alphas are ephemeral and their universe is very fluid. It takes quantitatively sophisticated, technologically well-endowed and ever-adapting trading operations to mine hundreds of thousands, millions and even billions of alphas and combine them into a unified "mega-alpha", which is then traded with an added bonus of sizeable savings on execution costs due to automatic internal crossing of trades.

---

[20] Suppressing alpha weights by the turnover can add value but be highly correlated with volatility suppression.

[21] Roughly speaking, when the specific risk is computed via nontrivial (proprietary) methods, the column in the factor loadings matrix corresponding to the turnover is no longer proportional to $\ln(\tau_i)$ but is a more complex function of the turnover, the specific risk also depends on the turnover nontrivially and is not quadratic in $\ln(\tau_i)$.

[22] For proprietary reasons, we are not at liberty to state precisely which ones.

In this spirit, we end this paper with an 1832 poem by a Russian poet Mikhail Lermontov (translation from Russian by Zura Kakushadze, ca. 1993):

*The Sail*

*A lonely sail seeming white,*
*In misty haze mid blue sea,*
*Be foreign gale seeking might?*
*Why home bays did it flee?*

*The sail's bending mast is creaking,*
*The wind and waves blast ahead,*
*It isn't happiness it's seeking,*
*Nor is it happiness it's fled!*

*Beneath are running ázure streams,*
*Above are shining golden beams,*
*But wishing storms the sail seems,*
*As if in storms is peace it deems.*

## Appendix A: Formulaic Alphas

In this appendix, in Subsection A.1, we provide our 101 formulaic alphas. The formulas are also code once the functions and operators are defined. The functions and operators used in the alphas are defined in Subsection A.2. The input data is elaborated upon in Subsection A.3.

### A.1. Formulaic Expressions for Alphas

Alpha#1: (rank(Ts_ArgMax(SignedPower(((returns < 0) ? stddev(returns, 20) : close), 2.), 5)) - 0.5)

Alpha#2: (-1 * correlation(rank(delta(log(volume), 2)), rank(((close - open) / open)), 6))

Alpha#3: (-1 * correlation(rank(open), rank(volume), 10))

Alpha#4: (-1 * Ts_Rank(rank(low), 9))

Alpha#5: (rank((open - (sum(vwap, 10) / 10))) * (-1 * abs(rank((close - vwap)))))

Alpha#6: (-1 * correlation(open, volume, 10))

Alpha#7: ((adv20 < volume) ? ((-1 * ts_rank(abs(delta(close, 7)), 60)) * sign(delta(close, 7))) : (-1 * 1))

Alpha#8: (-1 * rank(((sum(open, 5) * sum(returns, 5)) - delay((sum(open, 5) * sum(returns, 5)), 10))))

Alpha#9: ((0 < ts_min(delta(close, 1), 5)) ? delta(close, 1) : ((ts_max(delta(close, 1), 5) < 0) ? delta(close, 1) : (-1 * delta(close, 1))))

Alpha#10: rank(((0 < ts_min(delta(close, 1), 4)) ? delta(close, 1) : ((ts_max(delta(close, 1), 4) < 0) ? delta(close, 1) : (-1 * delta(close, 1)))))

Alpha#11: ((rank(ts_max((vwap - close), 3)) + rank(ts_min((vwap - close), 3))) * rank(delta(volume, 3)))

Alpha#12: (sign(delta(volume, 1)) * (-1 * delta(close, 1)))

Alpha#13: (-1 * rank(covariance(rank(close), rank(volume), 5)))

Alpha#14: ((-1 * rank(delta(returns, 3))) * correlation(open, volume, 10))

Alpha#15: (-1 * sum(rank(correlation(rank(high), rank(volume), 3)), 3))

Alpha#16: (-1 * rank(covariance(rank(high), rank(volume), 5)))

Alpha#17: (((-1 * rank(ts_rank(close, 10))) * rank(delta(delta(close, 1), 1))) * rank(ts_rank((volume / adv20), 5)))

Alpha#18: (-1 * rank(((stddev(abs((close - open)), 5) + (close - open)) + correlation(close, open, 10))))

Alpha#19: ((-1 * sign(((close - delay(close, 7)) + delta(close, 7)))) * (1 + rank((1 + sum(returns, 250)))))

Alpha#20: (((-1 * rank((open - delay(high, 1)))) * rank((open - delay(close, 1)))) * rank((open - delay(low, 1))))

Alpha#21: ((((sum(close, 8) / 8) + stddev(close, 8)) < (sum(close, 2) / 2)) ? (-1 * 1) : (((sum(close, 2) / 2) < ((sum(close, 8) / 8) - stddev(close, 8))) ? 1 : (((1 < (volume / adv20)) || ((volume / adv20) == 1)) ? 1 : (-1 * 1))))

Alpha#22: (-1 * (delta(correlation(high, volume, 5), 5) * rank(stddev(close, 20))))

Alpha#23: (((sum(high, 20) / 20) < high) ? (-1 * delta(high, 2)) : 0)

Alpha#24: ((((delta((sum(close, 100) / 100), 100) / delay(close, 100)) < 0.05) || ((delta((sum(close, 100) / 100), 100) / delay(close, 100)) == 0.05)) ? (-1 * (close - ts_min(close, 100))) : (-1 * delta(close, 3)))

Alpha#25: rank(((((-1 * returns) * adv20) * vwap) * (high - close)))

Alpha#26: (-1 * ts_max(correlation(ts_rank(volume, 5), ts_rank(high, 5), 5), 3))

Alpha#27: ((0.5 < rank((sum(correlation(rank(volume), rank(vwap), 6), 2) / 2.0))) ? (-1 * 1) : 1)

Alpha#28: scale(((correlation(adv20, low, 5) + ((high + low) / 2)) - close))

Alpha#29: (min(product(rank(rank(scale(log(sum(ts_min(rank(rank((-1 * rank(delta((close - 1), 5))))), 2), 1))))), 1), 5) + ts_rank(delay((-1 * returns), 6), 5))

Alpha#30: (((1.0 - rank(((sign((close - delay(close, 1))) + sign((delay(close, 1) - delay(close, 2)))) + sign((delay(close, 2) - delay(close, 3)))))) * sum(volume, 5)) / sum(volume, 20))

Alpha#31: ((rank(rank(rank(decay_linear((-1 * rank(rank(delta(close, 10)))), 10)))) + rank((-1 * delta(close, 3)))) + sign(scale(correlation(adv20, low, 12))))

Alpha#32: (scale(((sum(close, 7) / 7) - close)) + (20 * scale(correlation(vwap, delay(close, 5), 230))))

Alpha#33: rank((-1 * ((1 - (open / close))^1)))

Alpha#34: rank(((1 - rank((stddev(returns, 2) / stddev(returns, 5)))) + (1 - rank(delta(close, 1)))))

Alpha#35: ((Ts_Rank(volume, 32) * (1 - Ts_Rank(((close + high) - low), 16))) * (1 - Ts_Rank(returns, 32)))

Alpha#36: (((((2.21 * rank(correlation((close - open), delay(volume, 1), 15))) + (0.7 * rank((open - close)))) + (0.73 * rank(Ts_Rank(delay((-1 * returns), 6), 5)))) + rank(abs(correlation(vwap, adv20, 6)))) + (0.6 * rank((((sum(close, 200) / 200) - open) * (close - open)))))

Alpha#37: (rank(correlation(delay((open - close), 1), close, 200)) + rank((open - close)))

Alpha#38: ((-1 * rank(Ts_Rank(close, 10))) * rank((close / open)))

Alpha#39: ((-1 * rank((delta(close, 7) * (1 - rank(decay_linear((volume / adv20), 9)))))) * (1 + rank(sum(returns, 250))))

Alpha#40: ((-1 * rank(stddev(high, 10))) * correlation(high, volume, 10))

Alpha#41: (((high * low)^0.5) - vwap)

Alpha#42: (rank((vwap - close)) / rank((vwap + close)))

Alpha#43: (ts_rank((volume / adv20), 20) * ts_rank((-1 * delta(close, 7)), 8))

Alpha#44: (-1 * correlation(high, rank(volume), 5))

Alpha#45: (-1 * ((rank((sum(delay(close, 5), 20) / 20)) * correlation(close, volume, 2)) * rank(correlation(sum(close, 5), sum(close, 20), 2))))

Alpha#46: ((0.25 < (((delay(close, 20) - delay(close, 10)) / 10) - ((delay(close, 10) - close) / 10))) ? (-1 * 1) : (((((delay(close, 20) - delay(close, 10)) / 10) - ((delay(close, 10) - close) / 10)) < 0) ? 1 : ((-1 * 1) * (close - delay(close, 1)))))

Alpha#47: ((((rank((1 / close)) * volume) / adv20) * ((high * rank((high - close))) / (sum(high, 5) / 5))) - rank((vwap - delay(vwap, 5))))

Alpha#48: (indneutralize(((correlation(delta(close, 1), delta(delay(close, 1), 1), 250) * delta(close, 1)) / close), IndClass.subindustry) / sum(((delta(close, 1) / delay(close, 1))^2), 250))

Alpha#49: (((((delay(close, 20) - delay(close, 10)) / 10) - ((delay(close, 10) - close) / 10)) < (-1 * 0.1)) ? 1 : ((-1 * 1) * (close - delay(close, 1))))

Alpha#50: (-1 * ts_max(rank(correlation(rank(volume), rank(vwap), 5)), 5))

Alpha#51: (((((delay(close, 20) - delay(close, 10)) / 10) - ((delay(close, 10) - close) / 10)) < (-1 * 0.05)) ? 1 : ((-1 * 1) * (close - delay(close, 1))))

Alpha#52: ((((-1 * ts_min(low, 5)) + delay(ts_min(low, 5), 5)) * rank(((sum(returns, 240) - sum(returns, 20)) / 220))) * ts_rank(volume, 5))

Alpha#53: (-1 * delta((((close - low) - (high - close)) / (close - low)), 9))

Alpha#54: ((-1 * ((low - close) * (open^5))) / ((low - high) * (close^5)))

Alpha#55: (-1 * correlation(rank(((close - ts_min(low, 12)) / (ts_max(high, 12) - ts_min(low, 12)))), rank(volume), 6))

Alpha#56: (0 - (1 * (rank((sum(returns, 10) / sum(sum(returns, 2), 3))) * rank((returns * cap)))))

Alpha#57: (0 - (1 * ((close - vwap) / decay_linear(rank(ts_argmax(close, 30)), 2))))

Alpha#58: (-1 * Ts_Rank(decay_linear(correlation(IndNeutralize(vwap, IndClass.sector), volume, 3.92795), 7.89291), 5.50322))

Alpha#59: (-1 * Ts_Rank(decay_linear(correlation(IndNeutralize(((vwap * 0.728317) + (vwap * (1 - 0.728317))), IndClass.industry), volume, 4.25197), 16.2289), 8.19648))

Alpha#60: (0 - (1 * ((2 * scale(rank(((((close - low) - (high - close)) / (high - low)) * volume)))) - scale(rank(ts_argmax(close, 10))))))

Alpha#61: (rank((vwap - ts_min(vwap, 16.1219))) < rank(correlation(vwap, adv180, 17.9282)))

Alpha#62: ((rank(correlation(vwap, sum(adv20, 22.4101), 9.91009)) < rank(((rank(open) + rank(open)) < (rank(((high + low) / 2)) + rank(high))))) * -1)

Alpha#63: ((rank(decay_linear(delta(IndNeutralize(close, IndClass.industry), 2.25164), 8.22237)) - rank(decay_linear(correlation(((vwap * 0.318108) + (open * (1 - 0.318108))), sum(adv180, 37.2467), 13.557), 12.2883))) * -1)

Alpha#64: ((rank(correlation(sum(((open * 0.178404) + (low * (1 - 0.178404))), 12.7054), sum(adv120, 12.7054), 16.6208)) < rank(delta((((high + low) / 2) * 0.178404) + (vwap * (1 - 0.178404))), 3.69741))) * -1)

Alpha#65: ((rank(correlation(((open * 0.00817205) + (vwap * (1 - 0.00817205))), sum(adv60, 8.6911), 6.40374)) < rank((open - ts_min(open, 13.635)))) * -1)

Alpha#66: ((rank(decay_linear(delta(vwap, 3.51013), 7.23052)) + Ts_Rank(decay_linear(((((low * 0.96633) + (low * (1 - 0.96633))) - vwap) / (open - ((high + low) / 2))), 11.4157), 6.72611)) * -1)

Alpha#67: ((rank((high - ts_min(high, 2.14593)))^rank(correlation(IndNeutralize(vwap, IndClass.sector), IndNeutralize(adv20, IndClass.subindustry), 6.02936))) * -1)

Alpha#68: ((Ts_Rank(correlation(rank(high), rank(adv15), 8.91644), 13.9333) < rank(delta(((close * 0.518371) + (low * (1 - 0.518371))), 1.06157))) * -1)

Alpha#69: ((rank(ts_max(delta(IndNeutralize(vwap, IndClass.industry), 2.72412), 4.79344))^Ts_Rank(correlation(((close * 0.490655) + (vwap * (1 - 0.490655))), adv20, 4.92416), 9.0615)) * -1)

Alpha#70: ((rank(delta(vwap, 1.29456))^Ts_Rank(correlation(IndNeutralize(close, IndClass.industry), adv50, 17.8256), 17.9171)) * -1)

Alpha#71: max(Ts_Rank(decay_linear(correlation(Ts_Rank(close, 3.43976), Ts_Rank(adv180, 12.0647), 18.0175), 4.20501), 15.6948), Ts_Rank(decay_linear((rank(((low + open) - (vwap + vwap)))^2), 16.4662), 4.4388))

Alpha#72: (rank(decay_linear(correlation(((high + low) / 2), adv40, 8.93345), 10.1519)) / rank(decay_linear(correlation(Ts_Rank(vwap, 3.72469), Ts_Rank(volume, 18.5188), 6.86671), 2.95011)))

Alpha#73: (max(rank(decay_linear(delta(vwap, 4.72775), 2.91864)), Ts_Rank(decay_linear(((delta(((open * 0.147155) + (low * (1 - 0.147155))), 2.03608) / ((open * 0.147155) + (low * (1 - 0.147155)))) * -1), 3.33829), 16.7411)) * -1)

Alpha#74: ((rank(correlation(close, sum(adv30, 37.4843), 15.1365)) < rank(correlation(rank(((high * 0.0261661) + (vwap * (1 - 0.0261661)))), rank(volume), 11.4791))) * -1)

Alpha#75: (rank(correlation(vwap, volume, 4.24304)) < rank(correlation(rank(low), rank(adv50), 12.4413)))

Alpha#76: (max(rank(decay_linear(delta(vwap, 1.24383), 11.8259)), Ts_Rank(decay_linear(Ts_Rank(correlation(IndNeutralize(low, IndClass.sector), adv81, 8.14941), 19.569), 17.1543), 19.383)) * -1)

Alpha#77: min(rank(decay_linear(((((high + low) / 2) + high) - (vwap + high)), 20.0451)), rank(decay_linear(correlation(((high + low) / 2), adv40, 3.1614), 5.64125)))

Alpha#78: (rank(correlation(sum(((low * 0.352233) + (vwap * (1 - 0.352233))), 19.7428), sum(adv40, 19.7428), 6.83313))^rank(correlation(rank(vwap), rank(volume), 5.77492)))

Alpha#79: (rank(delta(IndNeutralize(((close * 0.60733) + (open * (1 - 0.60733))), IndClass.sector), 1.23438)) < rank(correlation(Ts_Rank(vwap, 3.60973), Ts_Rank(adv150, 9.18637), 14.6644)))

Alpha#80: ((rank(Sign(delta(IndNeutralize(((open * 0.868128) + (high * (1 - 0.868128))), IndClass.industry), 4.04545)))^Ts_Rank(correlation(high, adv10, 5.11456), 5.53756)) * -1)

Alpha#81: ((rank(Log(product(rank((rank(correlation(vwap, sum(adv10, 49.6054), 8.47743))^4)), 14.9655))) < rank(correlation(rank(vwap), rank(volume), 5.07914))) * -1)

Alpha#82: (min(rank(decay_linear(delta(open, 1.46063), 14.8717)), Ts_Rank(decay_linear(correlation(IndNeutralize(volume, IndClass.sector), ((open * 0.634196) + (open * (1 - 0.634196))), 17.4842), 6.92131), 13.4283)) * -1)

Alpha#83: ((rank(delay(((high - low) / (sum(close, 5) / 5)), 2)) * rank(rank(volume))) / (((high - low) / (sum(close, 5) / 5)) / (vwap - close)))

Alpha#84: SignedPower(Ts_Rank((vwap - ts_max(vwap, 15.3217)), 20.7127), delta(close, 4.96796))

Alpha#85: (rank(correlation(((high * 0.876703) + (close * (1 - 0.876703))), adv30, 9.61331))^rank(correlation(Ts_Rank(((high + low) / 2), 3.70596), Ts_Rank(volume, 10.1595), 7.11408)))

Alpha#86: ((Ts_Rank(correlation(close, sum(adv20, 14.7444), 6.00049), 20.4195) < rank(((open + close) - (vwap + open)))) * -1)

Alpha#87: (max(rank(decay_linear(delta(((close * 0.369701) + (vwap * (1 - 0.369701))), 1.91233), 2.65461)), Ts_Rank(decay_linear(abs(correlation(IndNeutralize(adv81, IndClass.industry), close, 13.4132)), 4.89768), 14.4535)) * -1)

Alpha#88: min(rank(decay_linear(((rank(open) + rank(low)) - (rank(high) + rank(close))), 8.06882)), Ts_Rank(decay_linear(correlation(Ts_Rank(close, 8.44728), Ts_Rank(adv60, 20.6966), 8.01266), 6.65053), 2.61957))

Alpha#89: (Ts_Rank(decay_linear(correlation(((low * 0.967285) + (low * (1 - 0.967285))), adv10, 6.94279), 5.51607), 3.79744) - Ts_Rank(decay_linear(delta(IndNeutralize(vwap, IndClass.industry), 3.48158), 10.1466), 15.3012))

Alpha#90: ((rank((close - ts_max(close, 4.66719)))^Ts_Rank(correlation(IndNeutralize(adv40, IndClass.subindustry), low, 5.38375), 3.21856)) * -1)

Alpha#91: ((Ts_Rank(decay_linear(decay_linear(correlation(IndNeutralize(close, IndClass.industry), volume, 9.74928), 16.398), 3.83219), 4.8667) - rank(decay_linear(correlation(vwap, adv30, 4.01303), 2.6809))) * -1)

Alpha#92: min(Ts_Rank(decay_linear(((((high + low) / 2) + close) < (low + open)), 14.7221), 18.8683), Ts_Rank(decay_linear(correlation(rank(low), rank(adv30), 7.58555), 6.94024), 6.80584))

Alpha#93: (Ts_Rank(decay_linear(correlation(IndNeutralize(vwap, IndClass.industry), adv81, 17.4193), 19.848), 7.54455) / rank(decay_linear(delta(((close * 0.524434) + (vwap * (1 - 0.524434))), 2.77377), 16.2664)))

Alpha#94: ((rank((vwap - ts_min(vwap, 11.5783)))^Ts_Rank(correlation(Ts_Rank(vwap, 19.6462), Ts_Rank(adv60, 4.02992), 18.0926), 2.70756)) * -1)

Alpha#95: (rank((open - ts_min(open, 12.4105))) < Ts_Rank((rank(correlation(sum(((high + low) / 2), 19.1351), sum(adv40, 19.1351), 12.8742))^5), 11.7584))

Alpha#96: (max(Ts_Rank(decay_linear(correlation(rank(vwap), rank(volume), 3.83878), 4.16783), 8.38151), Ts_Rank(decay_linear(Ts_ArgMax(correlation(Ts_Rank(close, 7.45404), Ts_Rank(adv60, 4.13242), 3.65459), 12.6556), 14.0365), 13.4143)) * -1)

Alpha#97: ((rank(decay_linear(delta(IndNeutralize(((low * 0.721001) + (vwap * (1 - 0.721001))), IndClass.industry), 3.3705), 20.4523)) - Ts_Rank(decay_linear(Ts_Rank(correlation(Ts_Rank(low, 7.87871), Ts_Rank(adv60, 17.255), 4.97547), 18.5925), 15.7152), 6.71659)) * -1)

Alpha#98: (rank(decay_linear(correlation(vwap, sum(adv5, 26.4719), 4.58418), 7.18088)) - rank(decay_linear(Ts_Rank(Ts_ArgMin(correlation(rank(open), rank(adv15), 20.8187), 8.62571), 6.95668), 8.07206)))

Alpha#99: ((rank(correlation(sum(((high + low) / 2), 19.8975), sum(adv60, 19.8975), 8.8136)) < rank(correlation(low, volume, 6.28259))) * -1)

Alpha#100: (0 - (1 * (((1.5 * scale(indneutralize(indneutralize(rank(((((close - low) - (high - close)) / (high - low)) * volume)), IndClass.subindustry), IndClass.subindustry))) - scale(indneutralize((correlation(close, rank(adv20), 5) - rank(ts_argmin(close, 30))), IndClass.subindustry))) * (volume / adv20))))

Alpha#101: ((close - open) / ((high - low) + .001))

## A.1. Functions and Operators

(Below "{ }" stands for a placeholder.  All expressions are case insensitive.)

abs(x), log(x), sign(x) = standard definitions; same for the operators "+", "-", "*", "/", ">", "<", "==", "||", "x ? y : z"

rank(x) = cross-sectional rank

delay(x, d) = value of x d days ago

correlation(x, y, d) = time-serial correlation of x and y for the past d days

covariance(x, y, d) = time-serial covariance of x and y for the past d days

scale(x, a) = rescaled x such that sum(abs(x)) = a (the default is a = 1)

delta(x, d) = today's value of x minus the value of x d days ago

signedpower(x, a) = x^a

decay_linear(x, d) = weighted moving average over the past d days with linearly decaying weights d, d – 1, ..., 1 (rescaled to sum up to 1)

indneutralize(x, g) = x cross-sectionally neutralized against groups g (subindustries, industries, sectors, etc.), i.e., x is cross-sectionally demeaned within each group g

ts_{O}(x, d) = operator O applied across the time-series for the past d days; non-integer number of days d is converted to floor(d)

ts_min(x, d) = time-series min over the past d days

ts_max(x, d) = time-series max over the past d days

ts_argmax(x, d) = which day ts_max(x, d) occurred on

ts_argmin(x, d) = which day ts_min(x, d) occurred on

ts_rank(x, d) = time-series rank in the past d days

min(x, d) = ts_min(x, d)

max(x, d) = ts_max(x, d)

sum(x, d) = time-series sum over the past d days

product(x, d) = time-series product over the past d days

stddev(x, d) = moving time-series standard deviation over the past d days

*A.2. Input Data*

returns = daily close-to-close returns

open, close, high, low, volume = standard definitions for daily price and volume data

vwap = daily volume-weighted average price

cap = market cap

adv{d} = average daily dollar volume for the past d days

IndClass = a generic placeholder for a binary industry classification such as GICS, BICS, NAICS, SIC, etc., in indneutralize(x, IndClass.level), where level = sector, industry, subindustry, etc. Multiple IndClass in the same alpha need not correspond to the same industry classification.

## Appendix B: Disclaimer

Wherever the context so requires, the masculine gender includes the feminine and/or neuter, and the singular form includes the plural and vice versa.  The authors of this paper ("Authors") and their affiliates including without limitation Quantigic® Solutions LLC ("Authors' Affiliates" or "their Affiliates") make no implied or express warranties or any other representations whatsoever, including without limitation implied warranties of merchantability and fitness for a particular purpose, in connection with or with regard to the content of this paper including without limitation any formulae, code or algorithms contained herein ("Content").

The reader may use the Content solely at his/her/its own risk and the reader shall have no claims whatsoever against the Authors or their Affiliates and the Authors and their Affiliates shall have no liability whatsoever to the reader or any third party whatsoever for any loss, expense, opportunity cost, damages or any other adverse effects whatsoever relating to or arising from the use of the Content by the reader including without any limitation whatsoever: any direct, indirect, incidental, special, consequential or any other damages incurred by the reader, however caused and under any theory of liability; any loss of profit (whether incurred directly or indirectly), any loss of goodwill or reputation, any loss of data suffered, cost of procurement of substitute goods or services, or any other tangible or intangible loss; any reliance placed by the reader on the completeness, accuracy or existence of the Content or any other effect of using the Content; and any and all other adversities or negative effects the reader might encounter in using the Content irrespective of whether the Authors or their Affiliates are or should have been aware of such adversities or negative effects.

The formulae and code included in Appendix A hereof are provided herein with the express permission of WorldQuant LLC. WorldQuant LLC retains all rights, title and interest in and to the formulae and code included in Appendix A hereof and any and all copyrights therefor.

# Tables

| Quantity | Minimum | 1st Quartile | Median | Mean | 3rd Quartile | Maximum |
|---|---|---|---|---|---|---|
| $S$ | 1.238 | 1.929 | 2.224 | 2.265 | 2.498 | 4.162 |
| $T$ | 0.1571 | 0.3429 | 0.4752 | 0.5456 | 0.6474 | 1.604 |
| $1/T$ | 0.6235 | 1.545 | 2.104 | 2.391 | 2.916 | 6.365 |
| $C$ | 0.1324 | 0.3125 | 0.3969 | 0.4814 | 0.5073 | 2.031 |
| $10^3 \times \sigma$ | 0.9318 | 1.194 | 1.395 | 1.747 | 2.019 | 10.44 |
| $100\% \times \tilde{R}$ | 3.285 | 4.4 | 5.441 | 6.015 | 6.296 | 28.72 |
| $100\% \times \Psi_{ij}$ | -15.09 | 7.457 | 14.31 | 15.86 | 22.91 | 87.33 |

**Table 1.** Summary (using the R function `summary()`) for the annualized Sharpe ratio $S_i$, daily turnover, $T_i$, average holding period $1/T_i$, cents-per-share $C_i$, daily return volatility $\sigma_i$, annualized average daily return $\tilde{R}_i$, and pair-wise correlations $\Psi_{ij}$ with $i > j$ (see Section 3). The performance figures are exclusive of any trading or transaction costs, price impact, etc.

|  | Estimate | Standard error | t-statistic | Overall |
|---|---|---|---|---|
| Intercept | -3.509 | 0.295 | -11.88 |  |
| $\ln(\sigma)$ | 0.761 | 0.046 | 16.65 |  |
| Mult./Adj. R-squared |  |  |  | 0.737 / 0.734 |
| F-statistic |  |  |  | 277.2 |

**Table 2.** Summary (using the R function `summary(lm())`) for the cross-sectional regression of $\ln(R)$ over $\ln(\sigma)$ with the intercept. See Subsection 3.1 for details. Also see Figure 2.

|  | Estimate | Standard error | t-statistic | Overall |
|---|---|---|---|---|
| Intercept | -3.435 | 0.324 | -10.60 |  |
| $\ln(\sigma)$ | 0.775 | 0.052 | 14.84 |  |
| $\ln(T)$ | -0.023 | 0.040 | -0.57 |  |
| Mult./Adj. R-squared |  |  |  | 0.738 / 0.732 |
| F-statistic |  |  |  | 137.8 |

**Table 3.** Summary for the cross-sectional regression of $\ln(R)$ over $\ln(\sigma)$ and $\ln(T)$ with the intercept. See Subsection 3.1 for details.

|  | Estimate | Standard error | t-statistic | Overall |
|---|---|---|---|---|
| Intercept | 0.1587 | 0.0017 | 95.18 |  |
| $y_a$ | 0.0067 | 0.0023 | 2.907 |  |
| $z_a$ | 0.0474 | 0.0063 | 7.537 |  |
| Mult./Adj. R-squared |  |  |  | 0.0127 / 0.0123 |
| F-statistic |  |  |  | 32.55 |

**Table 4.** Summary for the cross-sectional regression of $\Psi_a$ over $y_a$ and $z_a$ with the intercept. See Subsection 3.2 for details. Also see Figure 3.

|  | Estimate | Standard error | t-statistic | Overall |
|---|---|---|---|---|
| Intercept | -6.174 | 0.062 | -100.1 |  |
| $\ln(T)$ | 0.368 | 0.068 | 5.412 |  |
| Mult./Adj. R-squared |  |  |  | 0.228 / 0.221 |
| F-statistic |  |  |  | 29.29 |

**Table 5.** Summary for the cross-sectional regression of $\ln(\sigma)$ over $\ln(T)$ with the intercept. See Subsection 3.2 for details. Also see Figure 4.

## Figures

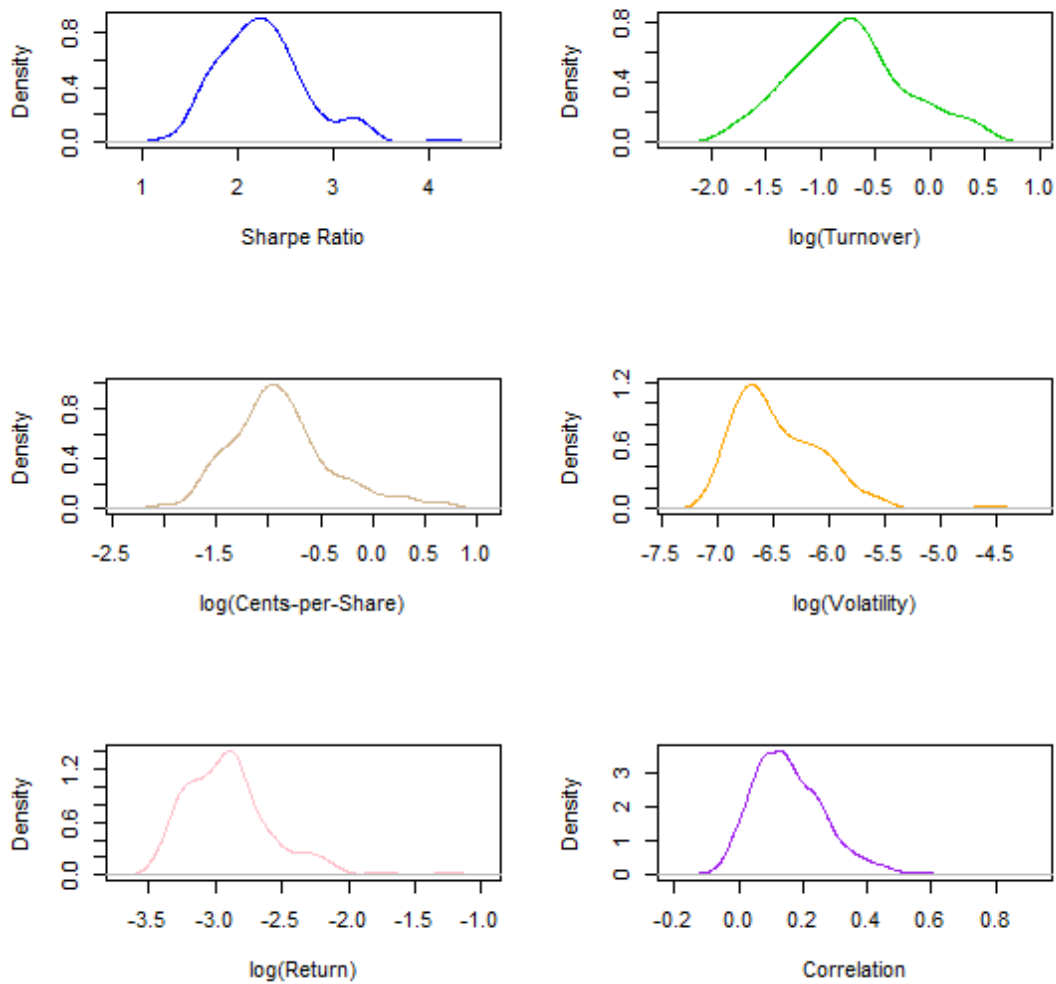

**Figure 1.** Density (using the R function `density()`) plots for the annualized Sharpe ratio $S_i$, daily turnover, $T_i$, cents-per-share $C_i$, daily return volatility $\sigma_i$, annualized average daily return $\tilde{R}_i$, and pair-wise correlations $\Psi_{ij}$ with $i > j$ (see Table 1 and Section 3). The "extreme" outliers in $S_i$, $\sigma_i$ and $\tilde{R}_i$ are due to the delay-0 alphas (see Section 2).

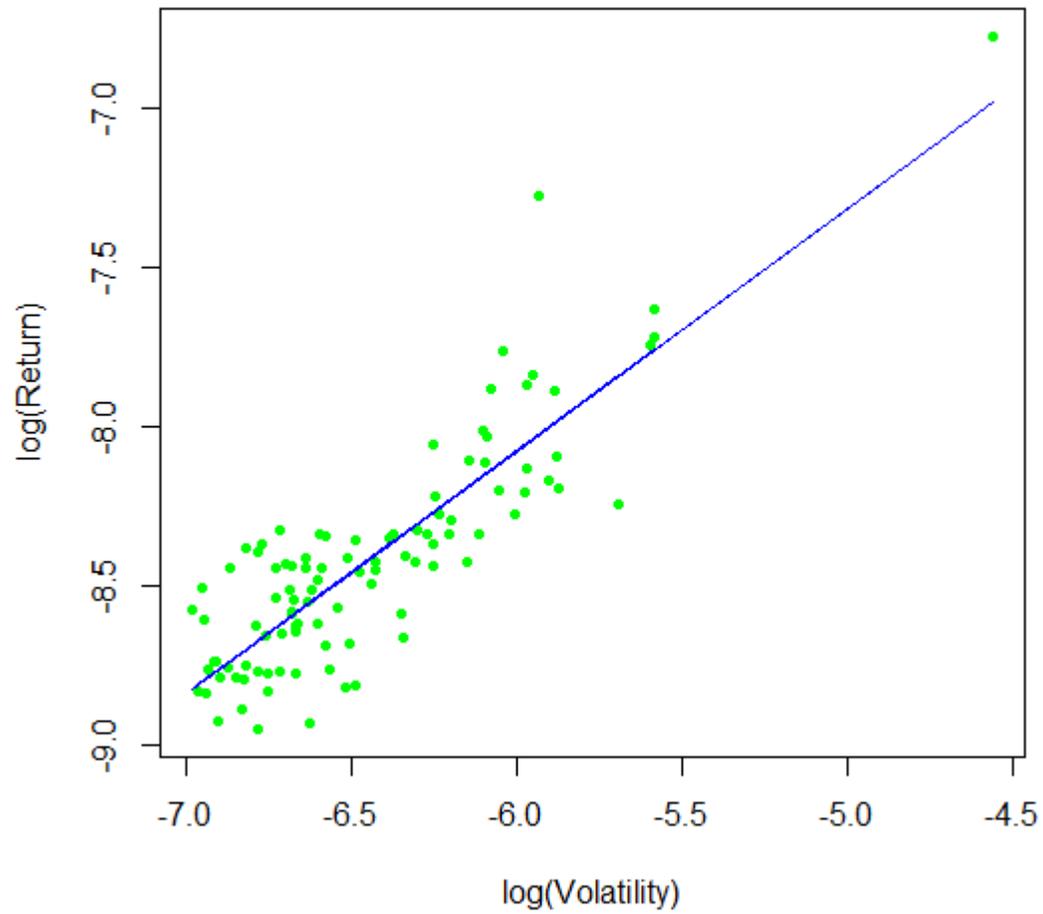

**Figure 2.** Horizontal axis: $\ln(\sigma)$; vertical axis: $\ln(R)$. The dots represent the data points. The straight line plots the linear regression fit $\ln(R) \approx -3.509 + 0.761 \ln(\sigma)$. See Table 2.

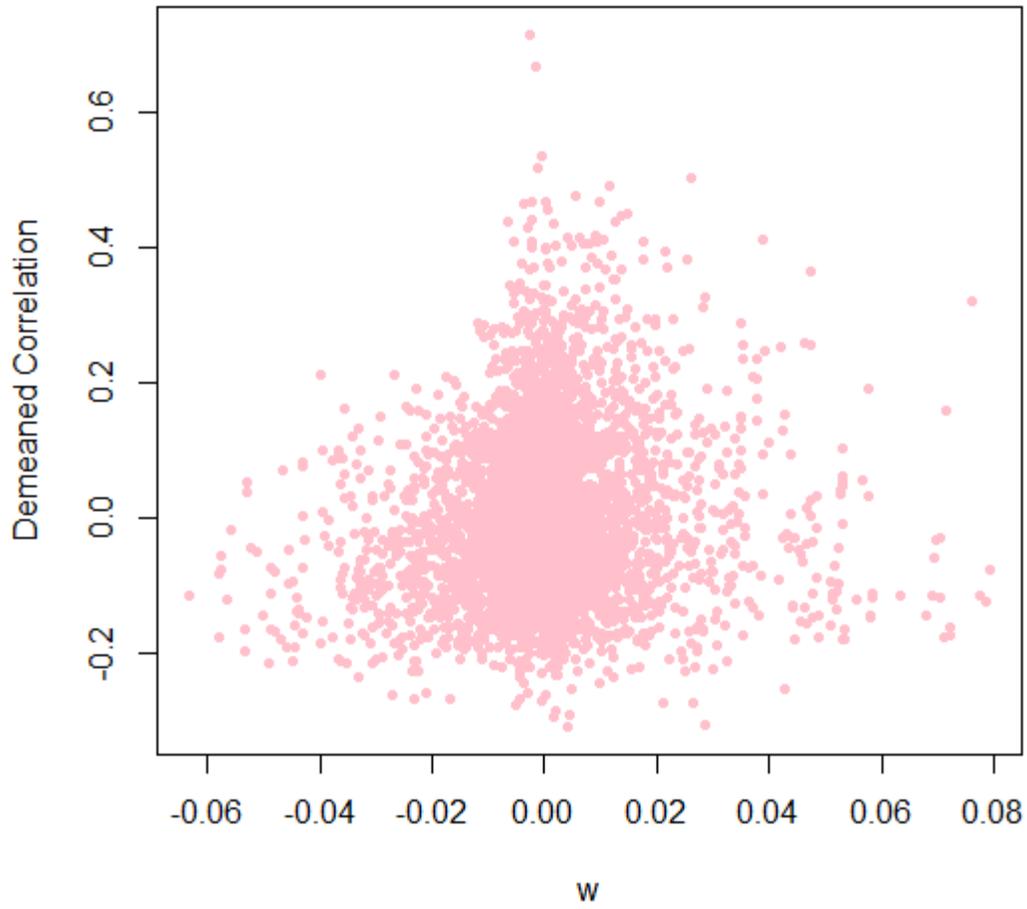

**Figure 3.** Horizontal axis: $w_a = 0.0067\, y_a + 0.0474\, z_a$; vertical axis: $\Psi_a - \text{Mean}(\Psi_a)$. See Table 4 and Subsection 3.2. The numeric coefficients are the regression coefficients in Table 4.

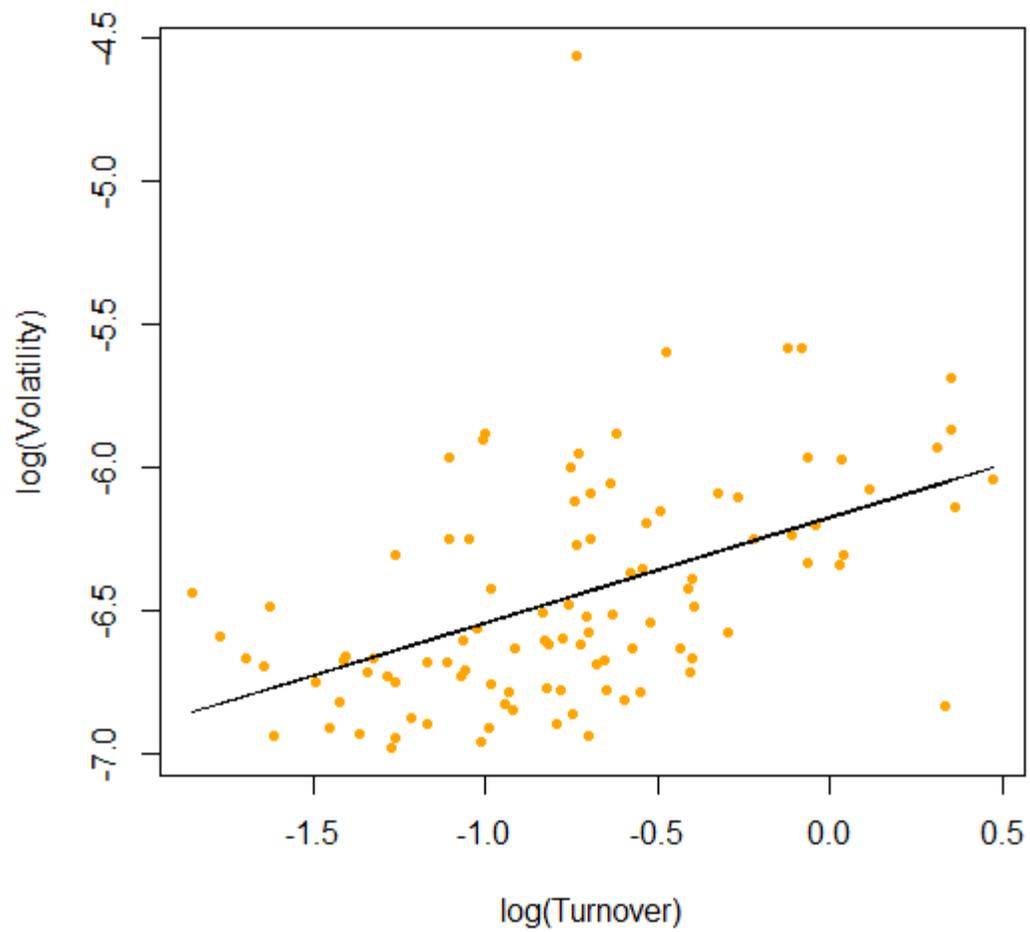

**Figure 4.** Horizontal axis: $\ln(T)$; vertical axis: $\ln(\sigma)$. The dots represent the data points. The straight line plots the linear regression fit $\ln(\sigma) \approx -6.174 + 0.368 \ln(T)$. See Table 5.